\documentclass[prl,twocolumn,superscriptaddress,showpacs,amsmath,amssymb]{revtex4-1}
\usepackage{graphicx}
\usepackage{dcolumn}
\usepackage{bm}
\usepackage{epstopdf}
\usepackage{upgreek}
\usepackage{color}
\newcommand{\blue}{\color{blue}}

\begin{document}

\title{Hydrogen-induced high-temperature superconductivity in two-dimensional materials: Example of hydrogenated monolayer MgB$_2$}

\author{J. Bekaert}
\email{jonas.bekaert@uantwerpen.be}
\affiliation{%
 Department of Physics, University of Antwerp,
 Groenenborgerlaan 171, B-2020 Antwerp, Belgium
}%
\author{M. Petrov}
\affiliation{%
 Department of Physics, University of Antwerp,
 Groenenborgerlaan 171, B-2020 Antwerp, Belgium
}%
\author{A. Aperis}
\affiliation{%
Department of Physics and Astronomy, Uppsala University,
Box 516, SE-751 20 Uppsala, Sweden
}
\author{P. M. Oppeneer}
\affiliation{%
Department of Physics and Astronomy, Uppsala University,
Box 516, SE-751 20 Uppsala, Sweden
}
\author{M. V. Milo\v{s}evi\'{c}}
\email{milorad.milosevic@uantwerpen.be}
\affiliation{%
 Department of Physics, University of Antwerp,
 Groenenborgerlaan 171, B-2020 Antwerp, Belgium
}

\date{\today}

\begin{abstract}
\noindent 
Hydrogen-based compounds under ultra-high pressure, such as the polyhydrides H$_3$S and LaH$_{10}$, superconduct through the conventional electron-phonon coupling mechanism to attain the record critical temperatures known to date. We demonstrate here that the intrinsic advantages of hydrogen for phonon-mediated superconductivity can be exploited in a completely different system, namely two-dimensional (2D) materials. We find that hydrogen adatoms can strongly enhance superconductivity in 2D materials due to flatband states originating from atomic-like hydrogen orbitals, with a resulting high density of states, and due to the emergence of high-frequency hydrogen-related phonon modes that boost the electron-phonon coupling. As a concrete example, we investigate the effect of hydrogen adatoms on the superconducting properties of monolayer MgB$_2$, by solving the fully anisotropic Eliashberg equations, in conjunction with a first-principles description of the electronic and vibrational states, and the coupling between them. We show that hydrogenation leads to a high critical temperature of 67 K, which can be boosted to over 100 K by biaxial tensile strain. 
\end{abstract}

\maketitle

\noindent In seminal work of 1968 Ashcroft showed that dense metallic hydrogen, if ever produced, could be a high-temperature superconductor \cite{PhysRevLett.21.1748}. The main reason would be its very high Debye temperature, as a result of its minimal mass, enabling very strong phonon-mediated superconducting pairing according to the Bardeen-Cooper-Schrieffer (BCS) theory. Subsequent detailed first-principles studies yielded critical temperature ($T_{\mathrm{c}}$) values up to 242 K, along with descriptions of the multiband nature of superconductivity \cite{PhysRevLett.100.257001}, and the role of phonon anharmonicities \cite{PhysRevLett.114.157004}. However, as creating metallic hydrogen requires immense pressures of $\sim$400 GPa \cite{PhysRevLett.112.165501,PhysRevLett.114.105305}, a confirmation of high-$T_{\mathrm{c}}$ superconductivity in pure hydrogen systems is still pending \cite{Diaseaal1579}.

Instead, in search of hydrogen-induced high-temperature superconductivity, most researchers have turned to \emph{polyhydrides}, compounds with a large hydrogen content, but also containing at least one other chemical element. The latter enables stabilizing the structure under lower applied pressure compared to metallic hydrogen itself. Notably, the chalcogen hydrides display experimentally proven high-$T_{\mathrm{c}}$ superconductivity, e.g., $T_{\mathrm{c}}=203$ K in H$_3$S, supplemented by comparable theoretical predictions for H-Te compounds \cite{PhysRevLett.116.057002}. The record $T_{\mathrm{c}}$'s among all currently known superconductors are held by the rare-earth hydrides, notably LaH$_{10}$, with $T_{\mathrm{c}}$ of $250-260$ K \cite{Drozdov2018Preprint,PhysRevLett.122.027001}, and there is also a theoretical prediction of an even higher $T_{\mathrm{c}}=303$ K in YH$_{10}$ \cite{PhysRevLett.119.107001}.

In this work we demonstrate a different approach to establish high-$T_{\mathrm{c}}$ superconductivity based on hydrogen, namely by adding hydrogen adatoms to two-dimensional (2D) superconductors \cite{Saito2016,0953-2048-30-1-013002,0953-2048-30-1-013003}, exploiting the changes in the electronic and vibrational properties that hydrogen induces. Such 2D superconductivity has been realized in recent years in very diverse ultrathin materials, ranging from atomically-thin elemental metal films (Pb, In, etc.) \cite{Zhang2010,NOFFSINGER2011421,PhysRevLett.110.237001}, over monolayers of unconventional superconductors \cite{Bollinger2011}, monolayer FeSe with high $T_{\mathrm{c}}$ up to 100 K \cite{PhysRevLett.112.107001,Ge2015}, to truly 2D atomic sheets such as doped graphene \cite{Kanetani27112012,Profeta2012,Ludbrook22092015,PhysRevB.94.064509} and transition metal dichalcogenides (TMDs) \cite{Xi2016,Ugeda2016,doi:10.1021/jacs.7b00216,PhysRevB.98.035203,Ye1193,Saito2015}. 

As an illustrative example, we consider hydrogenation of a monolayer (ML) magnesium diboride (MgB$_2$). Our main motivation to focus on this material is that it hosts distinct \emph{surface states} \cite{Bekaert2017b}, thus allowing for adatoms to be adsorbed at its surface. Further, ML MgB$_2$ structurally resembles lithium-doped graphene, another 2D superconductor of recent interest \cite{Profeta2012,Ludbrook22092015,PhysRevB.94.064509}. Finally, ML MgB$_2$ harbors three different types of electronic states at the Fermi level \cite{Bekaert2017} that may be prone to rich hybridization with adatoms, and a readily significant electron-phonon (\textit{e-ph}) coupling that may be boosted by vibrational contributions of the adatoms. As we will show, hydrogen indeed contributes to all those aspects of superconductivity in hydrogenated ML MgB$_2$ (H-MgB$_2$).

\begin{figure}[t]
\centering
\includegraphics[width=\linewidth]{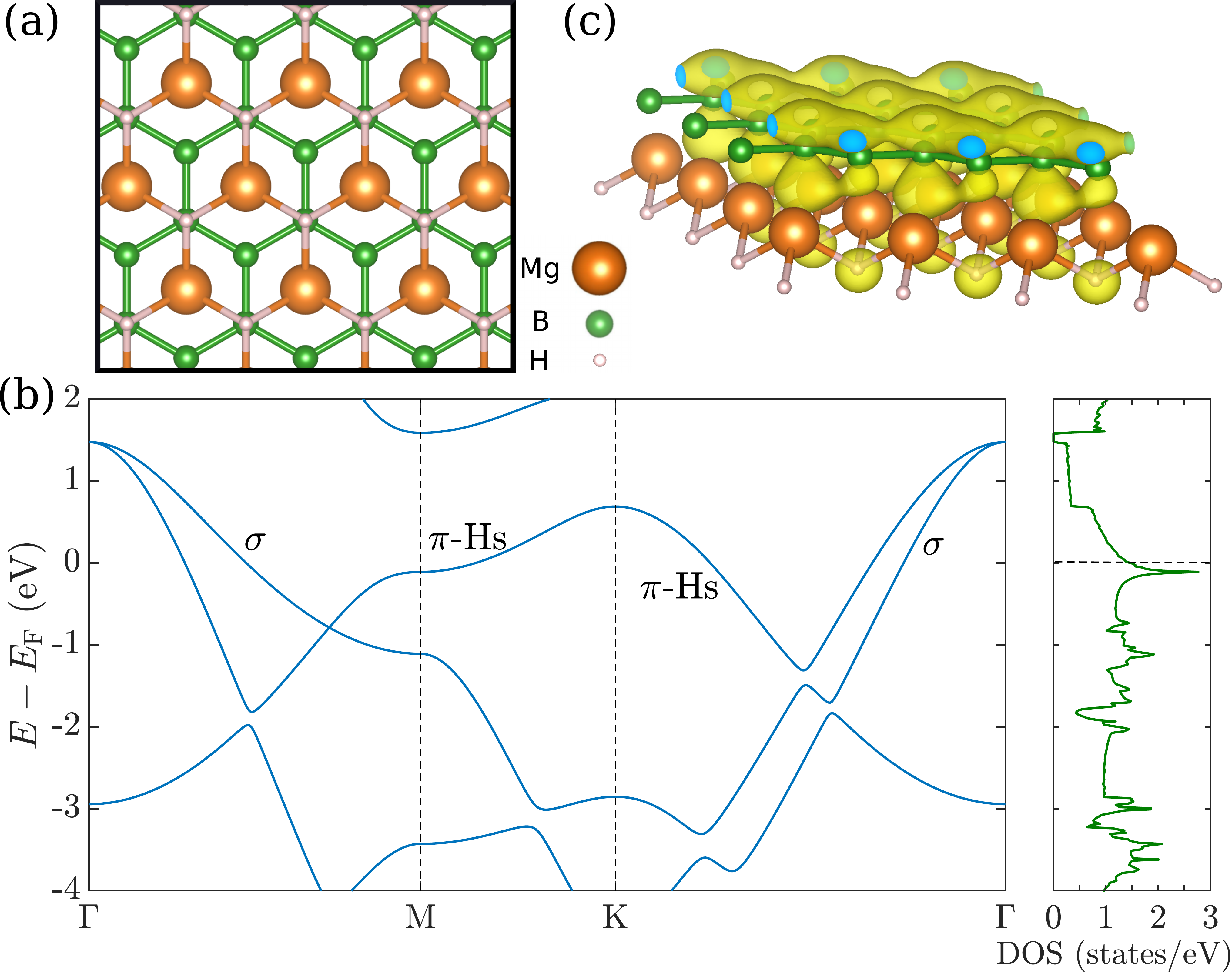}
\caption{(Color online) Structural and electronic properties of H-MgB$_2$. (a) Top view of the crystal structure. (b) The electronic band structure and density of states (DOS). The $\sigma$ and $\pi$-H{\blue \textit{s}} bands crossing $E_{\mathrm{F}}$ are also indicated. (c) The norm of the wave function of the $\pi$-H\textit{s} state at $E_{\mathrm{F}}$ obtained along the path $\Gamma$-K.}
\label{fig:fig1}
\end{figure}

\begin{figure*}[t]
\centering
\includegraphics[width=\linewidth]{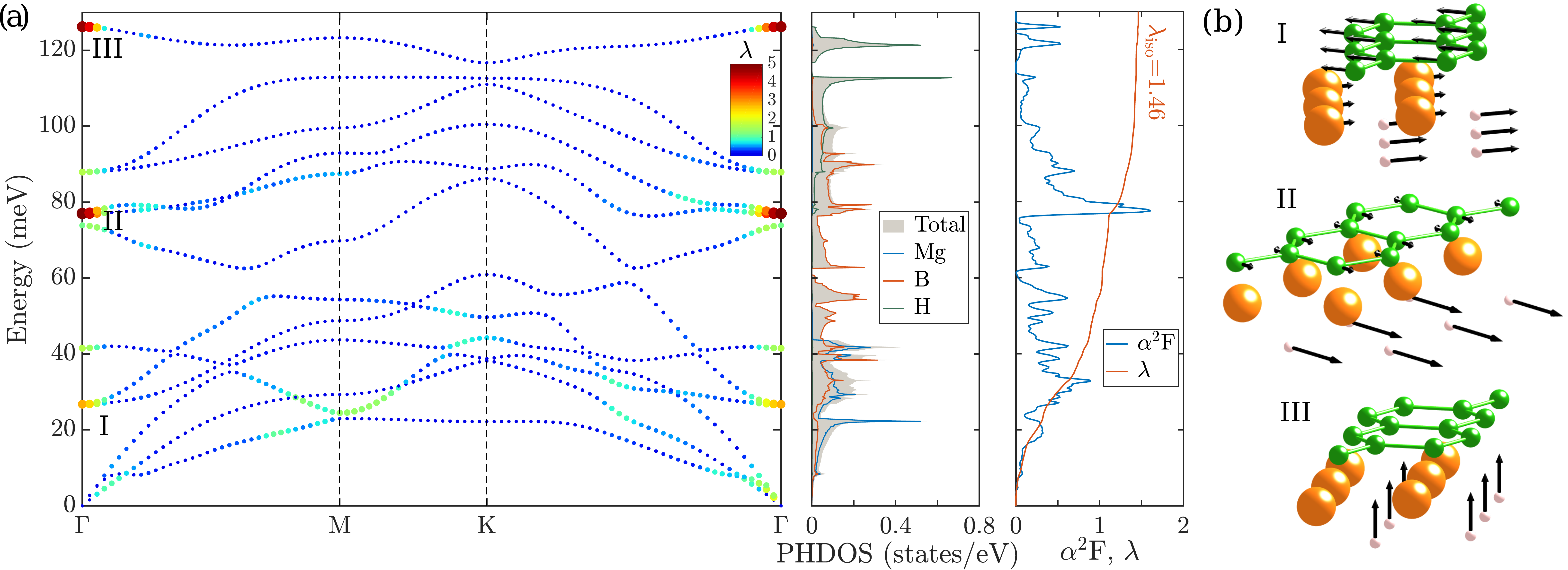}
\caption{(Color online) Phonons and electron-phonon interaction in H-MgB$_2$. (a) Phonon band structure (where both the colors and the dot sizes indicate the strength of the \textit{e-ph} coupling $\lambda$), phonon density of states (DOS, including the atom-resolved contributions), and isotropic Eliashberg function $\alpha^2F$ (plus the \textit{e-ph} coupling constant $\lambda$). (b) Phonon modes with the strongest \textit{e-ph} coupling, labelled as indicated in (a).}
\label{fig:fig2}
\end{figure*}

As the first step, we have determined the equilibrium structure of the H-MgB$_2$ system, based on density functional theory (DFT), as implemented in ABINIT \cite{Gonze20092582} [all computational details are provided in the Supplemental Material (SM)]. We found that the corresponding electronic structure depends critically on the concentration of hydrogen adatoms, as a result of the large charge transfer between the hydrogen and the magnesium layers. An optimal electronic structure for superconductivity is obtained in the case of one hydrogen atom per MgB$_2$ unit cell, resulting in the dynamically-stable structure shown in Fig.~\ref{fig:fig1}(a). The hydrogen adatoms are adsorbed on the magnesium side of the stack, with the same in-plane coordinate as one of the sublattices in boron's honeycomb lattice, resulting in a tricoordinated structure of the magnesium atoms. 

\begin{figure}[b]
\centering
\includegraphics[width=\linewidth]{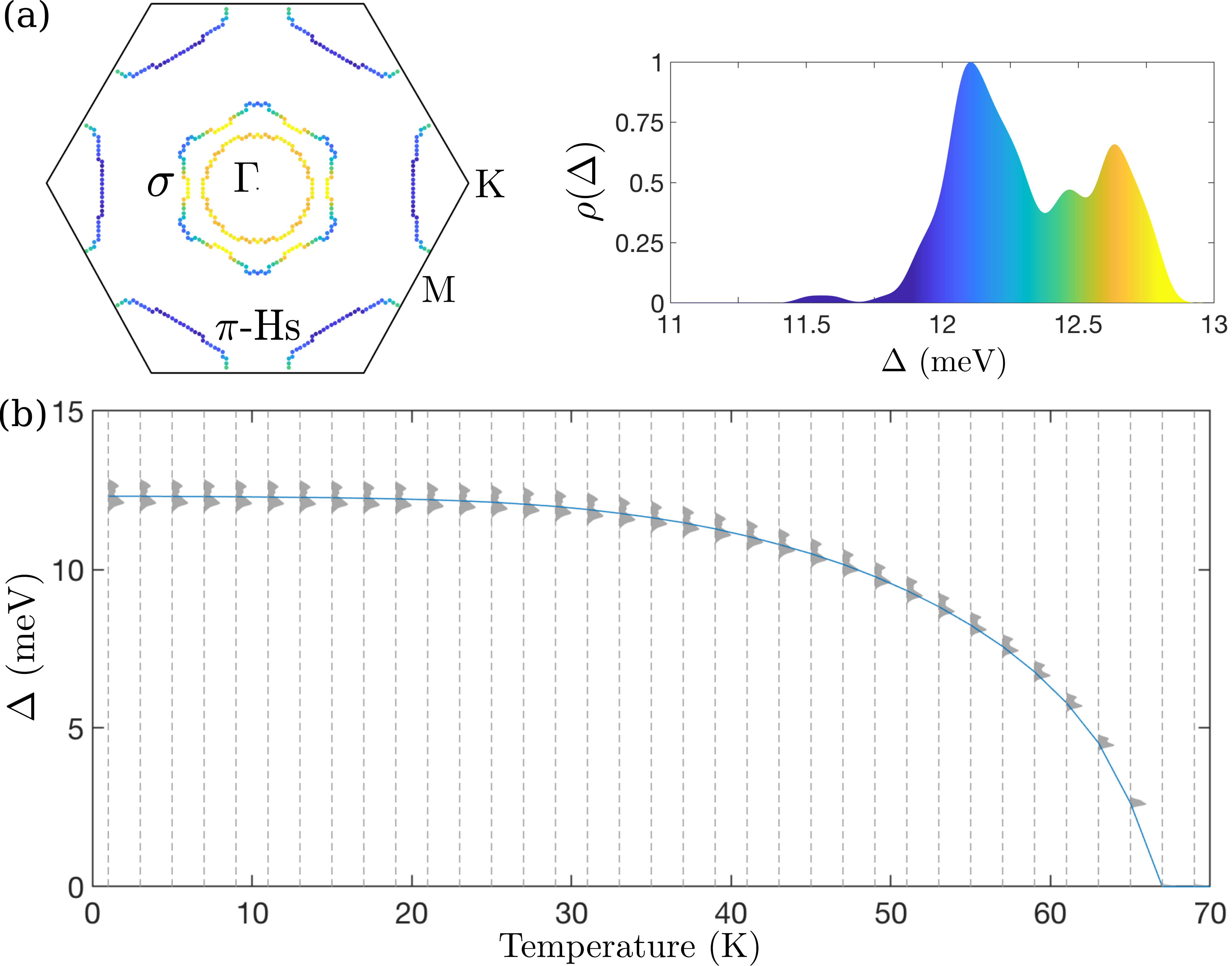}
\caption{(Color online) Superconducting properties of H-MgB$_2$, calculated with fully anisotropic Eliashberg theory. (a) The superconducting gap spectrum on the Fermi surface, calculated at 1 K, and the corresponding distribution $\rho(\Delta)$ with the color code. (b) The evolution of $\rho(\Delta)$ with temperature, yielding $T_{\mathrm{c}}=67$ K.}
\label{fig:fig3}
\end{figure}

The corresponding electronic band structure of H-MgB$_2$ is displayed in Fig.~\ref{fig:fig1}(b). Around the Fermi level ($E_{\mathrm{F}}$) the $\sigma$ band of ML MgB$_2$ (consisting of B-\textit{p}$_{x,y}$ states \cite{Bekaert2017}) persists, while the characteristic surface band stemming from the free Mg surface is eliminated by hydrogenation. Interestingly, the H-\textit{s} state forms a hybridized bond with the B-\textit{p}$_z$ states on the other side of the stack, as can be seen through the wave function in Fig.~\ref{fig:fig1}(c). Thus, a new electronic band emerges as a result of hydrogenation, indicated as the $\pi$-H\textit{s} state, in view of its hybridized band character. The resulting 2D Fermi surface consists of two $\sigma$ sheets around $\Gamma$ and the $\pi$-H\textit{s} state around K, as shown in Fig.~\ref{fig:fig3}(a). 

The hybridized $\pi$-H\textit{s} state in H-MgB$_2$ carries more density of states (DOS) than its pure $\pi$ counterpart in ML MgB$_2$, which is beneficial for superconductivity. The total DOS at $E_{\mathrm{F}}$ is $\sim 50\%$ larger in H-MgB$_2$ compared with ML MgB$_2$ (1.41 states/eV vs.\ 0.96 states/eV respectively). This DOS enhancement stems mainly from the flat dispersion of the $\pi$-H\textit{s} state around point M [cf.~Fig.~\ref{fig:fig1}(b)]. In this sense, the system studied here bears a resemblance to twisted bilayer graphene, where superconductivity appears due to a flatband interlayer state \cite{Cao2018}. We generally expect hydrogenation to contribute to a flatter dispersion due to the atomic-like H-\textit{s} state with very limited overlap [cf.~the inset in Fig.~\ref{fig:fig1}(b)]. Since the peak DOS value due to the flatband dispersion occurs merely 109 meV below $E_{\mathrm{F}}$, the superconducting properties of H-MgB$_2$ can be further (and significantly) enhanced with limited \textit{p}-doping so as to maximize the DOS at $E_{\mathrm{F}}$. 

As a next step, we determined the vibrational properties and the electron-phonon (\textit{e-ph}) coupling in H-MgB$_2$ using density functional perturbation theory (DFPT) calculations \cite{PhysRevLett.69.2819,RevModPhys.89.015003}. The resulting phonon band structure, shown in Fig.~\ref{fig:fig2}(a), proves the dynamical stability of the H-MgB$_2$ structure. In this phonon band structure, the mode($\nu$)- and momentum(\textbf{q})-dependent \textit{e-ph} coupling $\lambda^{\nu} (\textbf{q})$ is also indicated. The largest $\lambda^{\nu} (\textbf{q})$ values occur for $\textbf{q}=0$ (i.e.,~at $\Gamma$). The three modes with maximal coupling, depicted in Fig.~\ref{fig:fig2}(b), are (I) a shear mode of the B plane on one hand and the Mg and H planes on the other hand, (II) a B-B in-plane stretching mode where the H atoms move in-phase with the B atoms of the other sublattice, and (III) Mg and H moving out-of-phase in the out-of-plane direction. Note that in all three modes H atoms are involved. Mode II is similar to the strongly coupling C-C stretching mode in superconducting \textit{p}-doped graphane, where the H atoms move along \cite{PhysRevLett.105.037002}. Therefore, apart from the electronic hybridization described above, there is also a hybridization of the vibrational modes of H with those of the other atoms. In general H atoms are prone to vibrationally hybridize as a result of their minimal mass. This is corroborated by the phonon DOS shown in Fig.~\ref{fig:fig2}(a), where a significant overlap of the H and B contributions can be observed. 
Finally, in the utmost right panel of Fig.~\ref{fig:fig2}(a) the isotropically averaged Eliashberg spectral function $\alpha^2F$ \cite{ALLEN19831} of the frequency-dependent \textit{e-ph} coupling is displayed. The resulting isotropic \textit{e-ph} coupling constant attains a high value of $\lambda_{\mathrm{iso}}=1.46$ in ML H-MgB$_2$, compared with $\lambda_{\mathrm{iso}}=0.68$ in non-hydrogenated ML MgB$_2$ \cite{Bekaert2017}. This strong enhancement of $\lambda$ due to hydrogenation is a combined effect of a higher electronic DOS and emerging strong-coupling phonon modes, both stemming from hybridization with electronic and vibrational states of hydrogen, as discussed above.
 
To accurately calculate the influence of hydrogenation on the superconducting properties of ML H-MgB$_2$, we used our first-principles results for electrons, phonons and their coupling field as input to solve the fully anisotropic Migdal-Eliashberg equations as implemented in the Uppsala Superconductivity code (UppSC) \cite{PhysRevB.92.054516,PhysRevB.94.144506,Bekaert2017,Bekaert2017b,PhysRevB.97.014503,PhysRevB.97.060501} (see SM for computational details). Notwithstanding the high phonon frequencies of H-MgB$_2$ due to hydrogen [cf.~Fig.\ \ref{fig:fig2}(a)], Migdal's approximation \cite{ALLEN19831} is valid in H-MgB$_2$, as the electronic characteristic energy scale still dominates (the ratio of the maximum phonon energy  $E_{\mathrm{ph}}$ and the electronic bandwidth $E_{\mathrm{e}}$--limited by the $\pi$-H\textit{s} state--amounts to $E_{\mathrm{ph}}/E_{\mathrm{e}}=0.18 \ll 1$). 

By solving the fully anisotropic formalism we obtain the superconducting gap, $\Delta(\textbf{k})$, as a function of electron momentum \textbf{k} (specifically the Fermi wave vectors). The results shown in Fig.~\ref{fig:fig3}(a) indicate strong Cooper pairing ($\Delta/T_{\mathrm{c}}=2.24$ as opposed to the BCS ratio of 1.76) in ML H-MgB$_2$, with $\Delta$ values in the range $11.5-13$ meV (at a low temperature of 1 K). $\Delta$ in the range $11.5-12.2$ meV stems predominantly from the H-\textit{s} state, whereas the highest $\Delta$ values are due to the $\sigma$ bands. Nevertheless, the distribution $\rho(\Delta)$ does not present separated domes for each contributing state, i.e., multigap behavior, as is the case in ML MgB$_2$ \cite{Bekaert2017}. We attribute the mixing of the pairing strengths upon hydrogenation to the emergence of significant interband coupling, through $\lambda^{\nu} (\textbf{q})$ with non-zero $\textbf{q}$ [cf.~Fig.~\ref{fig:fig2}(a)]. Such interband scattering is particularly strong between the two $\sigma$ states and the $\pi$-H\textit{s} state (see SM for further information).

Fig.~\ref{fig:fig3}(b) shows the temperature evolution of $\rho(\Delta)$, obtained from solving the Eliashberg equations for each temperature separately. This gives $T_{\mathrm{c}}=67$ K, the temperature where $\Delta$ vanishes. Since ML MgB$_2$ has $T_{\mathrm{c}}=20$ K \cite{Bekaert2017}, hydrogenation has strongly enhanced the superconducting pairing strength and the corresponding $T_{\mathrm{c}}$. In order to assess the contributions of the different chemical elements in H-MgB$_2$, we also investigated the \emph{isotope effect} \footnote{The results shown in Fig.~\ref{fig:fig3}(a) have been obtained with weighted averages of the isotope masses, according to their natural occurrence.}. To this end we renormalized the phonon frequencies according to the isotope masses, and subsequently solved again the Eliashberg equations. Apart from the B isotope effect that is essentially unchanged compared to pure MgB$_2$ (see SM for details), we found a quite strong H isotope effect. Namely, $T_{\mathrm{c}}$ decreases from 67 K to 63 K when hydrogen is replaced with deuterium, and further down to 59 K for tritium, supporting again the crucial role of H in the high-temperature superconductivity of H-MgB$_2$. 

In all of these calculations we have described the Coulomb interaction between the Cooper-pair electrons using a Morel-Anderson pseudopotential \cite{PhysRev.125.1263} of $\mu^*=0.13$. We note however that  our main result, the strong enhancement of superconductivity due to hydrogenation, persists even for an increased Coulomb interaction, as detailed in the Supplemental Material. 

\begin{figure}[t]
\centering
\includegraphics[width=\linewidth]{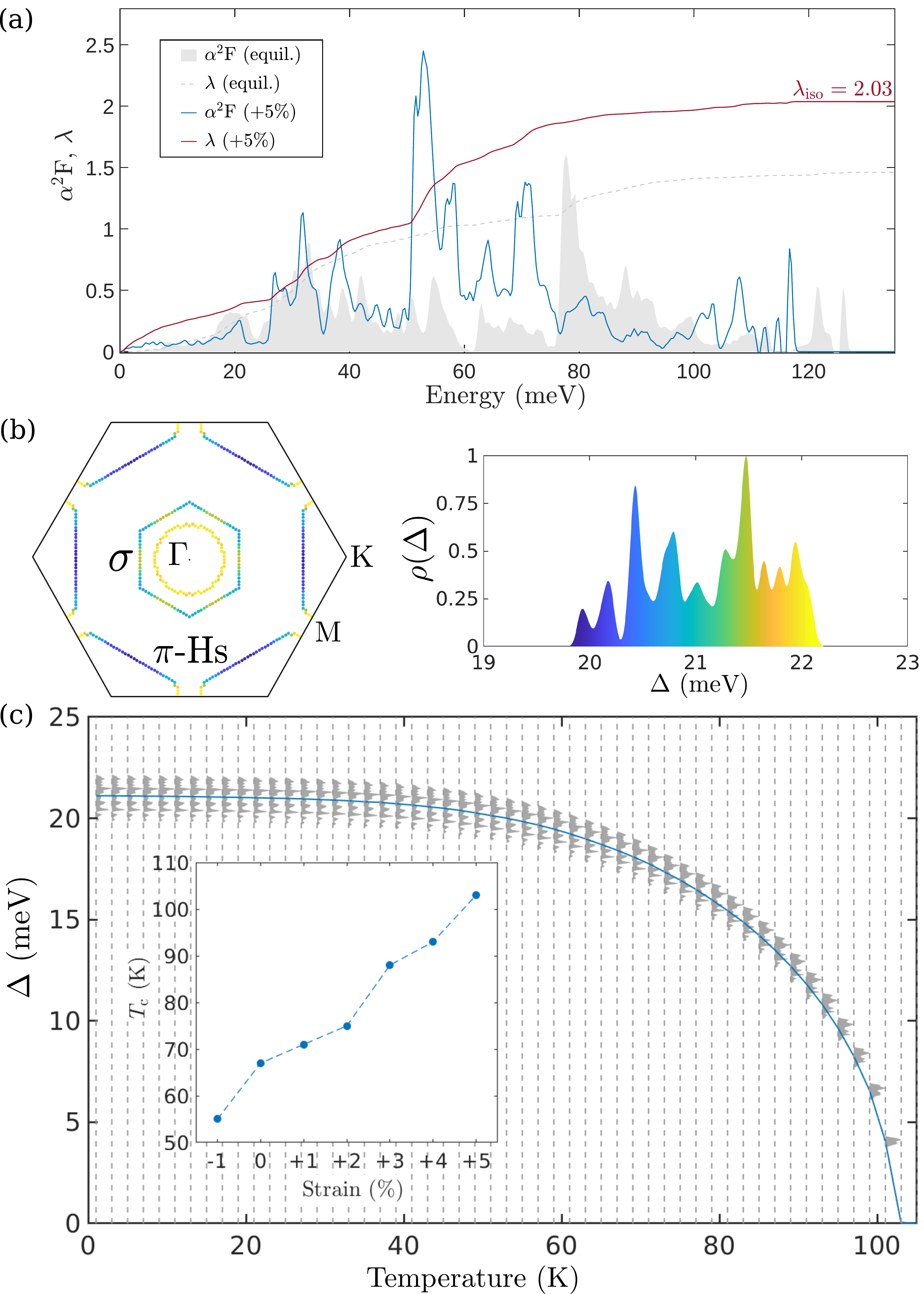}
\caption{(Color online) Superconducting properties of H-MgB$_2$ under the influence of strain. (a) Isotropic Eliashberg function, $\alpha^2F$, and electron-phonon coupling constant, $\lambda$, for applied 5\% tensile strain. The data for the non-strained case are shown by the grey shadow for comparison. (b) The superconducting gap spectrum on the Fermi surface, from fully anisotropic calculations for applied 5\% tensile strain, and the corresponding color-coded $\rho(\Delta)$. (c) The evolution of $\rho(\Delta)$ with temperature for 5\% tensile strain, yielding $T_{\mathrm{c}}=103$ K. The inset shows $T_{\mathrm{c}}$ as a function of strain.}
\label{fig:fig4}
\end{figure}

At this point, we recall that the properties of 2D materials are relatively easily manipulated by lattice deformation or strain, due to a substrate or by exerting a force on a suspended monolayer \cite{0295-5075-108-6-67005,PhysRevLett.111.196802,C5NR07755A,Bekaert2017}. To check the corresponding effect on superconductivity in H-MgB$_2$, we applied in-plane biaxial strain, where the hexagonal symmetry of the lattice is preserved. We carried out first-principles calculations with strain in the range -5\% (compressive) to 5\% (tensile) strain. We found that hydrogen greatly stiffens the material due to its intrinsically high phonon frequencies, so that H-MgB$_2$ cannot be compressively strained beyond -1\%, yet the phonon spectra proved stable for tensile strain up to 5\%. The change in the Eliashberg function when the in-plane lattice parameter is increased by 5\%, as calculated from first principles, is shown in Fig.~\ref{fig:fig4}(a). Due to such applied tensile strain the phonon frequencies lower overall, and in addition some of the main peak amplitudes increase. This increase stems from changes in the electronic structure, as with tensile strain the \emph{flatband peak} in the electronic DOS moves closer to $E_{\mathrm{F}}$, such that the Fermi sheets of the $\pi$-H\textit{s} band increase, as shown in Fig.~\ref{fig:fig4}(b) [cf.~Fig.~\ref{fig:fig3}(a)]. These combined changes in $\alpha^2F$ lead to a strong boost of the \textit{e-ph} coupling, to $\lambda_{\mathrm{iso}}=2.03$ (compared with 1.46 in the unstrained case), since $\lambda$ relates to the integral over frequencies of $\alpha^2F(\omega)/\omega$. 

On the base of the obtained first-principles results we carried out anisotropic Eliashberg calculations, to visualize the changes in the superconducting properties. The obtained gap spectrum $\Delta(\textbf{k})$ and corresponding distribution $\rho(\Delta)$ in case of 5\% tensile strain displayed in Fig.~\ref{fig:fig4}(b) reveal a great boost in $\Delta$ with respect to the unstrained case, to the range of $20-22$ meV, as a result of the enhanced \textit{e-ph} coupling. Such high $\Delta$ values are characteristic of high-temperature superconductivity. In Fig.~\ref{fig:fig4}(c) we show the temperature evolution of $\Delta$, from which we obtain $T_{\mathrm{c}}=103$ K. In the inset we show the $T_{\mathrm{c}}$ values for all studied strain values, yielding an almost linear dependence. By using substrates with a slightly larger lattice constant (e.g., Si$_{1+x}$C$_{1-x}$ or Al$_x$Ga$_{1-x}$N alloys, with a lattice constant tunable by $x$ \cite{PhysRevB.73.024509}) tensile strain can thus be applied to obtain desired $\Delta$ values and to increase $T_{\mathrm{c}}$ of H-MgB$_2$ above 100 K.

In conclusion, our study establishes hydrogen as an ideal adatomic candidate to enhance the superconducting properties of 2D materials, and to boost their critical temperatures to the high-$T_{\mathrm{c}}$ range. We showed that hydrogen adatoms adsorbed on a 2D superconductor open new electron-phonon coupling channels at high frequencies, and that hydrogen's electronic and vibrational states are very prone to hybridization, thereby enhancing pre-existing Cooper-pairing channels. We demonstrated that a major advantage of these hybridized electronic states is their \emph{flatband} character, owing to hydrogen's atomic-like \textit{s}-orbital. The high electronic density of states due to such flatband dispersion strongly enhances the electron-phonon coupling, akin to the flatband dispersion in case of twisted bilayer graphene \cite{Cao2018}. 

As a concrete example, we showed that the above reasons yield a high $T_{\mathrm{c}}$ of 67 K in a hydrogenated monolayer of magnesium diboride, that can be boosted above 100 K by merely 5\% of tensile biaxial strain. This proves that hydrogenation of a 2D material can indeed lead to strong electron-phonon coupling and high-$T_{\mathrm{c}}$ superconductivity, as exploited in bulk hydride compounds with record $T_{\mathrm{c}}$'s to date \cite{Drozdov2018Preprint,PhysRevLett.122.027001}, yet without the need to apply excessively high pressures that hamper practical applications.

\begin{acknowledgments}
This work was supported by TOPBOF-UAntwerp, Research Foundation-Flanders (FWO), the Swedish Research Council (VR), the R{\"o}ntgen-{\AA}ngstr{\"o}m Cluster and the EU-COST Action CA16218. J.B. acknowledges support as a postdoctoral fellow of the FWO. J.B. and M.P. contributed equally to this work. The computational resources and services used for the first-principles calculations in this work were provided by the VSC (Flemish Supercomputer Center), funded by the FWO and the Flemish Government -- department EWI. Eliashberg theory calculations were supported through the Swedish National Infrastructure for Computing (SNIC). 
\end{acknowledgments}

\bibliography{biblio}

\end{document}